\documentclass[11pt]{article}             
\title{Probability representation and state-extended uncertainty relations}
\author{V.~N.~Chernega$^{\dag}$, V.~I.~Man'ko$^{\dag}$
        \\$^{\dag}$ P.N.~Lebedev Physical Institute, Russian Academy of Sciences\\
        Leninskii Prospect, 53, Moscow 119991, Russia \\
        \\Emails: vchernega@gmail.com, manko@sci.lebedev.ru}

\begin{document}

\maketitle

\begin{abstract}
The new inequality recently found by Trifonov and called the
state-extended inequality is considered in the
tomographic-probability representation of quantum mechanics. The
Trifonov uncertainty relations are expressed in terms of optical
tomograms and can be checked in experiments on homodyne detection of
the photon states.
\end{abstract}


\section{Introduction}
Nonclassical states of photons \cite{Dodonov2002} are the subject of
investigation in homodyne detecting experiments. The uncertainty
relations~\cite{Heis27}, being the basic principles of quantum
mechanics, can be presented in different forms.
Recently~\cite{Mancini96} a new formulation of quantum mechanics,
called the tomographic-probability representation of quantum
mechanics, was introduced. The tomographic approach provides the
possibility to express all quantum postulates and equations for the
wave function and density matrix in the form where the fair
probability distributions are used instead of the density matrices
(wave functions). A new modification of the uncertainty relations
was found by Trifonov~\cite{Trifonov2000}. These relations have not
yet been checked experimentally. In this paper, we present the
Trifonov uncertainty relations in the tomographic form, which
nowadays is the most appropriate method for checking experimentally
the basic principles of quantum mechanics by homodyne detection of
the photon states.

\section{State-Extended Uncertainly Relations}
 The standard uncertainty relation of Heisenberg~\cite{Heis27}
for the position and momentum looks as follows:
\begin{eqnarray}\label{eq.1}
\sigma_{qq}\sigma_{pp}\geq {1}/{4}.
\end{eqnarray}
We take the Planck's constant $\hbar=1$. For the pure state
$|\Psi\rangle$, (\ref{eq.1}) reads
\begin{equation}\label{eq.2}
\sigma_{qq}=\mbox{Tr}\left(\hat{q}^2|\Psi\rangle\langle\Psi|\right)-
(\mbox{Tr}\left(\hat{q}|\Psi\rangle\langle\Psi|\right))^2,
\qquad\sigma_{pp}=\mbox{Tr}\left(\hat{p}^2|\Psi\rangle\langle\Psi|\right)
-\left(\mbox{Tr}\left(\hat{p}|\Psi\rangle\langle\Psi|\right)\right)^2,
\end{equation}
which are dispersions of the position and momentum, respectively.

Trifonov~\cite{Trifonov2000} proved that this relation can be
generalized for two different states $|\Psi_1\rangle$ and
$|\Psi_2\rangle$. This generalization, called the state-extended
uncertainty relation, is
\begin{eqnarray}
\frac{1}{2}\left[\mbox{Tr}\left(\hat{q}^2|\Psi_1\rangle\langle\Psi_1|\right)
-(\mbox{Tr}\left(\hat{q}|\Psi_1\rangle\langle\Psi_1|\right))^2\right]
\left[\mbox{Tr}\left(\hat{p}^2|\Psi_2\rangle\langle\Psi_2|\right)-
(\mbox{Tr}\left(\hat{p}|\Psi_2\rangle\langle\Psi_2|\right))^2\right]\nonumber\\
+\frac{1}{2}\left[\mbox{Tr}\left(\hat{q}^2|\Psi_2\rangle\langle\Psi_2|\right)
-(\mbox{Tr}\left(\hat{q}|\Psi_2\rangle\langle\Psi_2|\right))^2\right]
\left[\mbox{Tr}\left(\hat{p}^2|\Psi_1\rangle\langle\Psi_1|\right)-
(\mbox{Tr}\left(\hat{p}|\Psi_1\rangle\langle\Psi_1|\right))^2\right]\geq
\frac{1}{4}.\label{eq.3}
\end{eqnarray}
Inequality~(\ref{eq.3}) provides constraints for dispersions of the
position and momentum. The difference between the Trifonov
inequality~(\ref{eq.3}) and the Heisenberg uncertainty
relation~(\ref{eq.1}) consists in the fact that two states
$|\Psi_1\rangle$ and $|\Psi_2\rangle$ are involved in (\ref{eq.3}),
while in the Heisenberg inequality~(\ref{eq.1}), only one state
$|\Psi\rangle$ is taken into account. One can see that, if in
(\ref{eq.3}) we take $|\Psi_1\rangle=|\Psi_2\rangle$,
inequality~(\ref{eq.3}) provides the Heisenberg uncertainty relation
as a particular case of inequality~(\ref{eq.3}).

Now we pose the question --- is it possible to check experimentally
the state-extended uncertainly relations~(\ref{eq.3})? For such a
check, we suggest to use homodyne detection of the photon states. At
the output of this experiment, one obtains the so-called optical
tomogram, which contains complete information on the photon quantum
state.

\section{Optical Tomography of Photon States}
 In this section, we review the approach called optical
tomography of quantum states where the fair probability-distribution
function $w(X,\theta)$ is used as an alternative to the wave
function and density matrix~\cite{Mancini96,ManciniFP,Ibort,RitaFP}.
The probability-distribution function $w(X,\theta)$ is called the
optical tomogram of the quantum state. The argument $X$ of the
function is called the homodyne quadrature of the photon, and the
angle $\theta$ is called the local oscillator phase --- and this
parameter is controlled in experiments on homodyne measuring of the
quantum states~\cite{Raymer,Lvovski,Mlynek}. The optical tomogram is
related to the wave function $\Psi(y)$ in view of the
formula~\cite{Mendes}
\begin{eqnarray}\label{eq.4}
w(X,\theta)=\frac{1}{2\pi
|\sin\theta|}\left|\int\Psi(y)\exp\left(\frac{iy^2}{2\tan\theta}-\frac{iXy}{\sin\theta}\right)\,dy\right|^2.
\end{eqnarray}
For the local oscillator phase $\theta=0$, the optical tomogram
provides the probability distribution of the position, i.e.,
\begin{eqnarray}\label{eq.5}
w(X,0)=|\Psi(x)|^2.
\end{eqnarray}
For the local oscillator phase $\theta=\pi/2$, the optical tomogram
provides the probability distribution of the momentum, i.e.,
\begin{eqnarray}\label{eq.6}
w(X,\pi/2)=|\tilde{\Psi}(X)|^2,
\end{eqnarray}
where $\tilde{\Psi}(X)$ is the quantum-state wave function
$|\Psi\rangle$ in the momentum representation, i.e.,
\begin{eqnarray}\label{eq.7}
\tilde{\Psi}(X)=\frac{1}{\sqrt{2\pi}}\int\Psi(y)e^{iXy}\,dy.
\end{eqnarray}
Thus, one can obtain dispersions of the position and momentum if the
tomogram $w(X,\theta)$ is used as a probability distribution,
namely,
\begin{eqnarray}
&&\mbox{Tr}\left(\hat{q}^2|\Psi\rangle\langle\Psi|\right)=\int w(X,0)X^2\,dX,\label{eq.8}\\
&&\mbox{Tr}\left(\hat{p}^2|\Psi\rangle\langle\Psi|\right)=\int
w(X,\pi/2)X^2\,dX.\label{eq.9}
\end{eqnarray}
In view of (\ref{eq.8}) and (\ref{eq.9}), the Heisenberg uncertainty
relation can be rewritten in the tomographic form (see e.g.,
\cite{OlgaJRLR,Ibort,Porzio}) as follows:
\begin{eqnarray}\label{eq.10}
\left[\int w(X,0)X^2\,dX-\left(\int
w(X,0)X\,dX\right)^2\right]\left[\int w(X,\pi/2)X^2\,dX-\left(\int
w(X,\pi/2)X\,dX\right)^2\right]\geq \frac{1}{4}\,.\nonumber\\
\end{eqnarray}
Formula (\ref{eq.10}) is the tomographic form of the Heisenberg
inequality (\ref{eq.1}).

\section{Trifonov Inequality in the Tomographic Representation}
 Inequality (\ref{eq.3}) can be presented in the tomographic
form. For this, we express dispersions of the position and momentum
in the first state $|\Psi_1\rangle$ and the second state
$|\Psi_2\rangle$ in terms of optical tomograms $w_1(X,\theta)$ and
$w_2(X,\theta)$ of these states. The expressions for the position
dispersions read
\begin{eqnarray}\label{eq.11}
\langle\Psi_1|\hat{q}^2|\Psi_1\rangle-(\langle\Psi_1|\hat{q}|\Psi_1\rangle)^2=\int
w_1(X,0)X^2\,dX-\left(\int w_1(X,0)X\,dX\right)^2
\end{eqnarray}
and
\begin{eqnarray}\label{eq.12}
\langle\Psi_2|\hat{q}^2|\Psi_2\rangle-(\langle\Psi_2|\hat{q}|\Psi_2\rangle)^2=\int
w_2(X,0)X^2\,dX-\left(\int w_2(X,0)X\,dX\right)^2.
\end{eqnarray}
The dispersions of the momentum in the first and second states are
\begin{eqnarray}\label{eq.13}
\langle\Psi_1|\hat{p}^2|\Psi_1\rangle-(\langle\Psi_1|\hat{p}|\Psi_1\rangle)^2=\int
w_1(X,\pi/2)X^2\,dX-\left(\int w_1(X,\pi/2)X\,dX\right)^2
\end{eqnarray}
and
\begin{eqnarray}\label{eq.14}
\langle\Psi_2|\hat{p}^2|\Psi_2\rangle-(\langle\Psi_2|\hat{p}|\Psi_2\rangle)^2=\int
w_2(X,\pi/2)X^2\,dX-\left(\int w_2(X,\pi/2)X\,dX\right)^2.
\end{eqnarray}
We can introduce the dispersions in a rotated reference frame. In
fact, the rotated quadrature operator
\begin{eqnarray}\label{eq.15}
\hat{X}=\hat{q}\cos\theta+\hat{p}\sin\theta
\end{eqnarray}
has the physical meaning of the position in the rotated reference
frame for the local oscillator phase $\theta$ and the physical
meaning of the momentum in the rotated reference frame for the local
oscillator phase $\theta+\pi/2$. In view of this, we can write the
Trifonov inequality not only for the position and momentum in the
initial reference frame but obtain the inequality in the rotated
reference frame.

For this, we replace in (\ref{eq.11})--(\ref{eq.14}) the zero local
oscillator phase by the local oscillator phase $\theta$. As a
result, we arrive at the inequality
\begin{eqnarray}
\frac{1}{2}\left[\int w_1(X,\theta)X^2\,dX-\left(\int
w_1(X,\theta)X\,dX\right)^2\right]\nonumber\\
\times\left[\int w_2(X,\theta+\pi/2)X^2\,dX-\left(\int
w_2(X,\theta+\pi/2)X\,dX\right)^2\right]
\nonumber\\
+\frac{1}{2}\left[\int w_2(X,\theta)X^2\,dX-\left(\int
w_2(X,\theta)X\,dX\right)^2\right]\nonumber\\
\times\left[\int w_1(X,\theta+\pi/2)X^2\,dX-\left(\int
w_1(X,\theta+\pi/2)X\,dX\right)^2\right]\geq{1}/{4}.\label{eq.16}
\end{eqnarray}
This inequality at $\theta=0$  provides the Trifonov inequality in
the tomographic form. Inequality~(\ref{eq.16}) can be checked in
experiments where two optical tomograms for two different photon
states are measured. In fact, inequality~(\ref{eq.16}) provides a
criterium for checking the accuracy of the experiments with homodyne
detection of photon states, since for arbitrary different pairs of
optical tomograms this inequality must be fulfilled. In the
classical domain, this inequality can be violated.
Relation~(\ref{eq.16}) can be easily extended to tomograms
corresponding to mixed quantum states.

For the pure state, the tomogram $w_1(X,\mu,\nu)$ determined by the
optical tomogram $w_1(X,\theta)$, in view of the formula
\begin{eqnarray}
w_1(X,\mu,\nu)=\frac{1}{\sqrt{\mu^2+\nu^2}}\,w_1\left
(\frac{X}{\sqrt{\mu^2+\nu^2}},\arctan\left({\nu}/{\mu}\right)
\right),\label{eq.17}
\end{eqnarray}
satisfies the integral equality
\begin{eqnarray}
\frac{1}{2\pi} \int w_1(X,\mu,\nu)w_1(Y,\mu,\nu)e^{i(X-Y)}
\,dX\,dY\,d\mu\,d\nu=1.\label{eq.18}
\end{eqnarray}

The same relation is valid for the pure-state tomogram
$w_2(X,\theta)$. Inequality (\ref{eq.16}) is correct if tomograms
$w_1(X,\theta)$ and $w_2(X,\theta)$ correspond to mixed states. This
means that they are the convex sums of tomograms satisfying
equality~(\ref{eq.18}). For mixed states, the tomograms provide a
value of integral~(\ref{eq.18}) smaller than $1$. Thus, the
tomographic form of the Trifonov inequality can be checked
experimentally not only for the pure quantum states but also for
mixed quantum states.

\section{Conclusions}
 To conclude, we point out the main results of this work.

We presented the state-extended uncertainty relations for the
position and momentum in the tomographic form. This form contains
only experimental optical tomograms that can be obtained by homodyne
detection of the photon states. We formulated the state-extended
inequality for the position and momentum and also found the
generalized inequality that is valid for an arbitrary local
oscillator phase. We suggest to use the tomographic state-extended
uncertainty relation obtained as a criterion for checking the
accuracy of homodyne measurements of the photon quantum states. We
propose to make an extra analysis of the homodyne experiments such
as those performed in
\cite{Raymer,Lvovski,Mlynek,Porzio,Zavatta1,Zavatta2} to check the
state-extended inequality. One can easily write the multimode
state-extended uncertainty relations~\cite{Trifonov2002} in the
tomographic form, and we will do this in future papers.

\section*{Acknowledgments}
 V.N.C. and V.I.M. thank the Russian Foundation for Basic
Research for partial support under Projects Nos.~09-02-00142 and
10-02-00312.


\begin{thebibliography}{99}

\bibitem{Dodonov2002} V.~V.~Dodonov, {\sl Quantum Semiclass. Opt.}, {\bf 4},
R1 (20002).

\bibitem{Heis27} W.~Heisenberg, {\sl Z. Phys.}, {\bf 43}, 172 (1927).

\bibitem{Mancini96}
S.~Mancini, V.~I.~Man'ko, and P.~Tombesi, {\sl Phys. Lett. A}, {\bf
213}, 1 (1996).

\bibitem{Trifonov2000}
D.~A.~Trifonov, {\sl J. Phys. A: Math. Gen.}, {\bf 33}, L299 (2000).

\bibitem{ManciniFP}
S. Mancini, V. I. Man'ko, and P. Tombesi, {\sl Found. Phys.}, {\bf
27}, 801 (1997).

\bibitem{Ibort} A. Ibort, V. I. Man'ko, G. Marmo, et al.,
{\sl Phys. Scr.}, {\bf 79}, 065013 (2009).

\bibitem{RitaFP}
M. A. Man'ko and V. I. Man'ko, {\sl Found. Phys.}, {\bf 41}, 330
(2011).

\bibitem{Raymer}
D. T. Smithey, M. Beck, M. G. Raymer, and A. Faridani, {\sl Phys.
Rev. Lett.}, {\bf 70}, 1244 (1993).

\bibitem{Lvovski}
A. I. Lvovsky and M. G. Raymer, {\sl Rev. Mod. Phys.}, {\bf 81}, 299
(2009).

\bibitem{Mlynek}
G. Breitenbach, S. Schiller, and J. Mlynek, {\sl Nature}, {\bf 387},
471 (1997).

\bibitem{Mendes}
V. I. Man'ko and R. V. Mendes, ArXiv Physica/9712022 Data Analysis,
Statistics, and Probability (1997); {\sl Phys. Lett. A}, {\bf 263},
53 (1999).

\bibitem{OlgaJRLR}
O. V. Man'ko and V. I. Man'ko, {\sl J. Russ. Laser Res.}, {\bf 25},
477 (2004).

\bibitem{Porzio} V. I. Man'ko, G. Marmo, A. Porzio, et al.,
``Homodyne estimation of quantum-states' purity by exploiting the
covariant uncertainty relation," ArXiv:quant-ph~1012.3297v1 (2010);
{\sl Phys. Scr.}, {\bf 83}, 045001 (2011).

\bibitem{Zavatta1}
V. Parigi, A. Zavatta, M. S. Kim, and M. Bellini, {\sl Science},
{\bf 317}, 1890 (2007).

\bibitem{Zavatta2}
A. Zavatta, S. Vicinai, and M. Bellini, {\sl Science}, \textbf{306},
660 (2004).

\bibitem{Trifonov2002}
D.~A.~Trifonov, {\sl Eur. Phys. J. B - Cond. Matter Complex Syst.},
{\bf 29}, 349 (2002).

\end{thebibliography}
\end{document}